\documentclass[conference,letterpaper,10pt]{IEEEtran}
\ifCLASSINFOpdf
\else
\fi
\usepackage{amsmath}
\usepackage{booktabs}

\usepackage[colorinlistoftodos]{todonotes}
\usepackage{graphicx}
\usepackage[strings]{underscore}
\let\labelindent\relax
\usepackage{enumitem}
\usepackage{cite}
\usepackage{verbatim}
\usepackage{multicol, blindtext}

\usepackage{xspace} 
\newcommand{\cnc}[0]{\textsc{C\&C}\xspace}

\newif\ifsubmit
\begin{document}
\title{Detection of Malicious and Low Throughput Data Exfiltration Over the DNS Protocol}

\ifsubmit
\author{Anonymous submission to Computers \& Security}
\else
\author{\IEEEauthorblockN{Asaf Nadler}
	\IEEEauthorblockA{Ben-Gurion University of the Negev\\
		and Akamai Technologies\\
		asafnadler@gmail.com}
	\and
	\IEEEauthorblockN{Avi Aminov}
	\IEEEauthorblockA{Akamai Technologies\\
		aviaminov@gmail.com}
	\and
	\IEEEauthorblockN{Asaf Shabtai}
	\IEEEauthorblockA{Ben-Gurion University of the Negev\\
		shabtaia@bgumail.bgu.ac.il}
}
\fi

\IEEEoverridecommandlockouts
\makeatletter\def\@IEEEpubidpullup{9\baselineskip}\makeatother
%\IEEEpubid{\parbox{\columnwidth}{}\hspace{\columnsep}\makebox[\columnwidth]{}}

\maketitle

\begin{abstract}

In the presence of security countermeasures, a malware designed for data exfiltration must use a covert channel to achieve its goal. The Domain Name System (DNS) protocol is a covert channel commonly used by malware developers today for this purpose. Although the detection of covert channels using the DNS has been studied for the past decade, prior research has largely dealt with a specific subclass of covert channels, namely DNS tunneling. While the importance of tunneling detection should not be minimized, an entire class of low throughput DNS exfiltration malware has been overlooked.

In this study we propose a method for detecting both tunneling and low throughput data exfiltration over the DNS. After determining that previously detected malware used Internet domains that were registered towards a cyber-campaign rather than compromising existing legitimate ones, we focus on detecting and denying requests to these domains as an effective data leakage shutdown. Therefore, our proposed solution handles streaming DNS traffic in order to detect and automatically deny requests to domains that are used for data exchange. The initial data collection phase collects DNS logs per domain in a manner that permits scanning for long periods of time, and is thus capable of dealing with ``low and slow'' attacks. The second phase extracts features based on the querying behavior of each domain, and in the last phase an anomaly detection model is used to classify domains based on their use for data exfiltration. As for detection, DNS requests to domains that were classified as being used for data exfiltration will be denied indefinitely.

Our method was evaluated on a large-scale recursive DNS server's logs with a peaking high of 47 million requests per hour. Within these DNS logs, we injected data exfiltration traffic from DNS tunneling tools as well as two real-life malware: FrameworkPOS, previously used for the theft of 56M credit cards from Home Depot in 2014, and Backdoor.Win32.Denis, which was active in the Cobalt Kitty APT in 2016. Even when restricting our method to an extremely low false positive rate (i.e., one in fifty thousand domains), it detected all of the above. In addition, the logs are used to compare our system with two recently published methods that focus on detecting DNS tunneling in order to stress the novelty of detecting low throughput exfiltration malware.

\end{abstract}

\ifsubmit
\else
\fi

% \fi
%\IEEEpeerreviewmaketitle

\section{Introduction} \label{sec:introduction}
Personal computers and computer networks have been the targets of data theft attacks commonly using techniques involving man-in-the-middle attacks~\cite{conti2016survey} or a malware that leaks data over a covert channel~\cite{zander2007survey, mccormick2008data}. In the case of a malware, usually, a remote server, which acts as a command and control (\cnc), waits for communications from the malware and logs the data transferred to it. However, in protected networks (private or organizational) the targeted hosts can reside in a restricted segment with limited access to/from the outside world. In such cases, even if connections are allowed, they are typically monitored by security solutions for suspicious behavior. Therefore, in such cases it is necessary for the malware to find a covert channel for the exfiltration of data to the remote server that will not be blocked or detected by the existing security solutions. One channel used for achieving this goal is the Domain Name System (DNS) protocol which is the focus of this study.

Any communication from local computers to the Internet (excluding static-IP based communications) relies on the DNS service. For that reason, restricting DNS communication may result in the disconnection of legitimate remote services; therefore, the DNS must be conservatively used to block content. From an attacker's perspective, this makes the DNS protocol a good candidate as a covert communication channel for data leakage~\cite{bromberger2011dns}.

Typically, the DNS protocol was not designed for arbitrary data exchange; DNS messages are relatively short and responses are uncorrelated which means that they do not necessarily arrive in the same order as the corresponding requests were sent~\cite{rfc1034}. These limitations can be dealt with by an attacker using one of two approaches: (1) establishing a bidirectional communication channel on top of the DNS (e.g., by emulating a reliable session between the malware and the \cnc server), or (2) using the channel to send small data points with minimal overhead and requests that are short and independent (e.g., credit card numbers, user credentials, key-logging and geographical locations). We refer to these two approaches as \emph{high throughput tunneling} and \emph{low throughput malware}, respectively. The complete threat landscape of DNS exfiltration is comprised of these two classes, and therefore, any solution for detecting or preventing data leakage over the DNS protocol must deal with both of them.

The problem of detecting covert channels has been widely studied in the last decade, both on a general level~\cite{gilbert2009approach,davis2016automated}, and on the DNS level~\cite{wang2016combating,homem2016entropy,cambiaso2016feature,buczak2016detection,engelstad2017detection,kara2014detection,sheridan2015detection}. However, to the best of our knowledge, previous work has mainly only focused on detecting the high throughput DNS tunneling communication class. While the case of tunneling is important, pursuing its unique characteristics, i.e., bi-directional data exchange and high throughput, may limit the ability of the proposed methods to detect the class of low throughput malware communication. Our survey indicates that there are at least ten known malware families using DNS exfiltration that have been overlooked entirely. When used in recent attacks, such low throughput exfiltration malware enabled credit card theft~\cite{krebs2014sally}, the control of compromised machines~\cite{shulmin2017tunn4comm}, and the installation of other malware~\cite{brumaghin2017dnsmessenger}, and thus their importance must not be underestimated.

Low throughput exfiltration malware detected in recent years used Internet domains that were purchased, registered, and operated solely for the sake of their cyber campaign (see subsection~\ref{subsec:landscape}). This is also the case for DNS tunneling tools in which a user is required to provide and configure an Internet domain. Therefore, denying requests to these domains is equivalent to stopping the data leakage without affecting normal network operation.

In this work we propose a novel DNS exfiltration detection approach based on machine learning techniques, which targets both DNS tunneling and low throughput malware exfiltration. The input for our approach is a stream of DNS traffic logs that are constantly grouped by domain. The logs are collected over a sufficiently long period of time to allow the detection of malware even if its data exchange rate is relatively slow. Periodically, a feature extraction phase is applied to the collected logs of each domain. Afterwards, a classification phase is performed on the features of each domain to determine if the domain is used for data exchange. The classification is performed using the Isolation Forest~\cite{liu2008isolation} anomaly detection model which is trained on legitimate traffic beforehand. Immediately after the classification, requests to any domain that has been classified as anomalous (i.e., used for data-exchange) will be blocked indefinitely, thus stopping the data leakage.

We evaluate our work using both tunneling tools and traffic simulations of previously detected low throughput exfiltration malware. These simulations are performed alongside legitimate large-scale DNS traffic with a peaking rate of 41 million requests per hour, thereby making the detection task as realistic as possible. After establishing detection coverage against these test subjects with less than 0.002\% false positive rates as well as a low increase in false positives over time, we compare our proposed method to two recently published papers aimed at detecting tunneling, in order to demonstrate our method's novel contribution --- detecting low throughput exfiltration malware.

The contribution of our work is three fold:
\begin{enumerate}
	\item In addition to the detection of DNS tunneling, our work is capable of detecting low throughput DNS exfiltration malware.
	\item Because our method classifies domains within a specific time frame (referred to as per domain), it performs the immediate, accurate, and automated blocking of emerging DNS data leakage attempts by denying requests to a domain that was classified as having been used for data exchange. Arguably, this is not the case for detection systems that classify users within a specific time frame (referred to as per user), since indefinite blocking of a legitimate user's traffic based on the existence of a malware might be unacceptable.
	\item To the best our knowledge, prior research has not been evaluated on large-scale, high quality DNS traffic logs (i.e., more than $10^{6}$ queries/hour). Since data exfiltration over the DNS may attempt to operate under the radar, testing against large-scale traffic is possibly the ultimate challenge for any detection system. The former claim is supported as when comparing our proposed method to previously suggested methods that were not necessarily designed for large-scale traffic, we see a reduction in their reported results.
\end{enumerate}

The rest of the paper is structured as follows. Section \ref{sec:background} contains a description of DNS exfiltration to allow greater understanding of our solution and its analysis. Section \ref{sec:related-works} presents the current state of research in the field. Our proposed method and its evaluation are presented in Sections \ref{sec:method} and \ref{sec:evaluation} respectively. The paper concludes with a discussion (Section \ref{sec:discussion}) and our conclusions and plans for future work (Section \ref{sec:summary-future-work}).

\section{Background} \label{sec:background}
In this section we briefly describe the DNS protocol~(\ref{subsec:protocl}), how it can be abused for data exchange (\ref{subsec:scheme}), how the abuse landscape includes both malicious and legitimate usage (\ref{subsec:landscape}), and how per domain classification instead of per user or per series of packets (referred to as per time) can be used to block malicious use (\ref{subsec:blocking}). In the last subsection we discuss the challenge of uncovering low throughput malware traffic (\ref{subsec:low throughput}).

\subsection{The DNS Protocol} \label{subsec:protocl}
The DNS protocol is a naming system for host machines and an essential component in the functionality of the Internet. The vast number of domains and subdomains on the Internet today exceeds the storage capabilities of a small, simple database. This was foreseen by the designers of the DNS, and the system was designed as a hierarchical distributed database. The resolution of a domain name to the IP address of its host machine starts by querying the root DNS servers (i.e. the head of the hierarchy) in a top-down manner until reaching a designated server called an \emph{authoritative name server} (AuthNS).

The authoritative name servers assigned to a domain name are responsible for resolving all of its descendants (i.e., subdomains). For example, if the login.example.com domain is queried, the DNS resolver will look for the AuthNS of example.com by querying the .com suffix. It will then query the AuthNS of example.com for the domain login.example.com and receive the answer (if it exists). While allowing scalability, the decentralization of the DNS allows any user to  configure the authoritative name server for their own domain names. Such a configuration is the foundation of the potential abuse of the DNS protocol towards the exchange of data.

\subsection{Data Exchange over the DNS Protocol} \label{subsec:scheme}
The DNS is designed as a stateless protocol for exchanging very short and specific types of information. At no point was this protocol meant to carry information from the client to the server interactively. However, the AuthNS sees all queries from the clients to the domains delegated to it, and if the queries follow a certain pattern, the requested subdomain can be interpreted as data. For example, if the passw0rd.example.com domain was queried, the AuthNS for example.com effectively receives the "passw0rd" string and can interpret it as incoming data.

\begin{figure} [ht]
	\centering
	\includegraphics[width=\columnwidth]{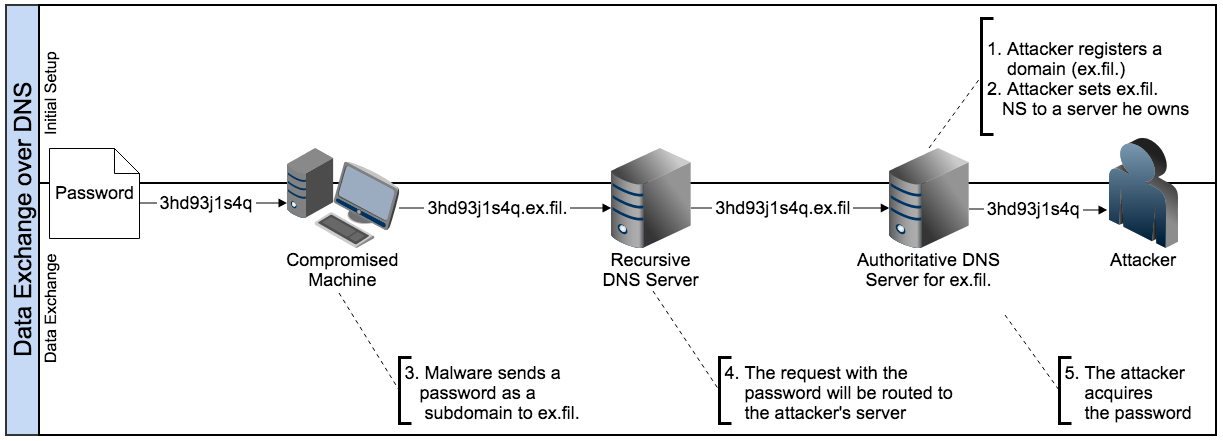}
	\caption{A general scheme for data exchange over the DNS protocol. The specific example portrays an attack in which a user password hosted on a compromised machine is unwillingly sent to an attacker via a DNS query.}
	\label{fig:scheme}
\end{figure}

In addition to the ability to send data to the AuthNS as described above, one can also leverage the response to form a bidirectional interactive data channel (see general scheme in Figure~\ref{fig:scheme}). Therefore, for the low cost of purchasing a domain and configuring a server as its AuthNS, an attacker can abuse the DNS, enabling it to serve as a de facto interactive communication channel between a querying machine and the server.

This type of abuse of the DNS protocol for the sake of data exchange has been thoroughly studied in previous research along with its unique attributes~\cite{homem2016entropy,kara2014detection,sheridan2015detection}. The most commonly investigated unique attributes are: long queries and responses~\cite{sheridan2015detection}, different resource record distribution~\cite{kara2014detection}, and a high volume of requests and encoded data rather than plain text~\cite{homem2016entropy}. While these abnormalities may capture the entire landscape of data exchange over the DNS, they are insufficient for accurate data leakage detection, since not all data exchange is malicious. 

\subsection{The Landscape of Data Exchange over the DNS} \label{subsec:landscape}
The scheme presented in subsection \ref{subsec:scheme} allows us to define the domains that are used for data exchange over the DNS. Formally, a domain for which any subquery contains a data payload to be delivered to its authoritative name server is considered a domain used for data exchange over the DNS. For example, the sophosxl.net domain. (Sophos Extensible List~\footnote{https://community.sophos.com/kb/en-us/31563}) has sub queries containing reversed IP addresses in the form of 1.1.168.192.ip.07.s.sophosxl.net. which are designed to be extracted by Sophos' authoritative name servers and responded to with an IP address if they appear in Sophos IP reputation systems. The fact that this method of operation is directed towards anti spam shows that data exchange over the DNS is not necessarily malicious.

Indeed, the landscape of domain names contains both legitimate and malicious domains. While the sophosxl.net domain mentioned above is a legitimate example, the 29a.de domain was formerly registered and used by BernhardPOS, a point of sale malware, in order to deliver stolen credit cards in the form of: PzMnPiosOD4nOCwuOzomPS4nN- jovPS8uOzsnNCstODkjOCwoMwAA.29a.de. The existence of both malicious and legitimate primary domains with regards to data exchange over the DNS may impose a distinguishing challenge if a domain belongs to both groups. 

Fortunately, this challenge has not yet been met. Previously detected malware used Internet domains that were registered and operated solely for the purpose of data leakage and were never used legitimately (see Table \ref{tbl:malware_c2}). The set of Internet domains used by these malware is comprised of either short domain names, to allow maximal bandwidth under the 255 byte per query limitation (e.g., 29a.de.), or the false imitation of known brands, to allow better ``hiding in plain sight'' (e.g., a23-33-37-54-deploy-akamaitechnlogies.com. is using the Combsquatting~\cite{kintis2017hiding} scheme and does not belong to Akamai Technologies, Inc.). Given that any domain is either malicious or benign but not both, subsection~\ref{subsec:blocking} explains how detected malicious domains can be blocked.

\begin{table}[htb!]
	\centering
	\caption{DNS Exfiltration Malware}
	\label{tbl:malware_c2}
	\begin{tabular}{@{}llll@{}}
			\toprule
			Year & Malware Name        & Targets & Domains Used \\
			\midrule
			2011 & Morto~\cite{mullaney2011morto}               & RDP / RAT & \begin{tabular}{@{}c@{}}ms[.]jifr[.]co[.]cc, \\ ms[.]jifr[.]net~\cite{symantec2011morto}\end{tabular}  \\
			2011 & FeederBot~\cite{dietrich2011feederbot}           & Botnet    & images[.]moviedyear[.]net~\cite{dietrich2016feederbot} \\
			2014 & PlugX~\cite{perigaud2014plugx}               & RDP / RAT & ns4[.]msftncsl[.]com~\cite{vasilenko2013plugx} \\
			2014 & FrameworkPOS~\cite{rascagneres2016new}              & POS       & \begin{tabular}{@{}c@{}}a23-33-37-54-deploy- \\ akamaitechnologies[.]com~\cite{medleta2016frameworkpos}\end{tabular}  \\
			2015 & Wekby~\cite{grunzweigh2016wekby}   & Targeted  & ns1[.]logitech-usa[.]com~\cite{grunzweigh2016wekby}  \\
			2015 & BernhardPOS~\cite{morphick2015bernhardpos}               & POS       & 29a[.]de~\cite{morphick2015bernhardpos} \\
			2015 & JAKU~\cite{forcepoint2015jaku} & Botnet    & LS4[.]com~\cite{forcepoint2015jaku} \\
			2016 & MULTIGRAIN~\cite{lync2016multigrain} & POS       & dojfgj[.]com~\cite{pandas2016multigrain} \\
			2017 & DNSMessenger~\cite{brumaghin2017dnsmessenger}        & Targeted  & cspg[.]pw, algew[.]me~\cite{brumaghin2017dnsmessenger}  \\
			\bottomrule	
	\end{tabular}\par
	\bigskip
The domains in use by previously detected malware.
\end{table}

\subsection{Blocking Malicious Data Exchange} \label{subsec:blocking}
Although the existence of seven known malware over the course of a decade may not lead one to believe that there is a vast amount of data leakage taking place over the DNS, the damage caused by these attacks cannot be minimized. Therefore, any data leakage detection system is measured on its ability to report data leakage that has occurred, as well as block emerging data leakage.

The automated blocking of a domain upon detection can be implemented by denying DNS requests to any descendant of that domain (i.e., returning a non-existent or server failure response) or by not answering at all. Blocking by denying requests is a common feature on most network security products (e.g., firewalls, secure DNS resolvers) which is largely enabled by a simple record change. This ease of acting on a threat right after its detection is not as straightforward for other DNS data leakage detection systems that act upon specific user traffic or per time, as the consistent use of legitimate services over the DNS may make it over-sensitive (e.g., queries to *.sophosxl.net at time $t$ will trigger an alert). Given that, we suggest that a per domain classification approach exceeds per user or per time approaches with regard to detection and blocking accuracy.

\subsection{Low Throughput Data Exchange} \label{subsec:low throughput}
After confirming the potential of blocking malicious domains based on traffic abnormality, we focus on the time frame in which the traffic is observed. For the case of high throughput DNS tunneling tools (see Table \ref{tbl:tunneling}), observing up to an hour of DNS traffic may suffice. However, for low throughput malware exfiltration, this may not be sufficient. 

\begin{table}[ht]
	\caption{High throughput tunneling tools}
	\centering
	\label{tbl:tunneling}
	\begin{tabular}{@{}lll@{}}
		\toprule
		Year & Tunnel Tool  & Platform\\
		\midrule
		2000 & NSTX & Linux							\\
		2004 & OzymanDNS & Linux (Perl)              \\
		2006 & Iodine Protocol v5.02 & Linux, Windows, Mac OS X  \\
		2008 & TCP-Over-DNS & Linux, Windows, Solaris  \\
		2008 & Split-brain & Linux                      \\
		2009 & Dns2tcp & Linux, Windows                \\
		2009 & Heyoka & Windows                         \\
		2010 & DNScat2 & Linux, Windows                \\
		2012 & MagicTunnel & Android					\\
		2012 & Element53 & Android						\\
		2015 & YourFreedom & Linux, Windows, Mac OS X  \\
		2016 & VPN-Over-DNS & Android					\\
		\bottomrule
	\end{tabular}\par
	\footnotesize{A list of known DNS tunneling tools}
\end{table}

Non-interactive data leakage channels (e.g., one-way credit card sending) may contain shorter messages and use a slower rate compared to a two-way interactive channel. Therefore, addressing this threat by inspecting a short time-window of traffic (e.g., an hour) might be insufficient for a designed detection algorithm. Instead, we suggest that any data leakage detection system should facilitate the efficient scanning of recent traffic with as few memory and disk limitations as possible. One way to tackle this problem is by using a sliding window approach, so that data is not recollected frequently.

A combination of the sliding window approach and the per domain approach can promote data collection efficiency (i.e., each log is collected once), and the detection range can be increased as mush as possible given the system's memory and disk usage. This makes the system capable of detecting even the ``lowest and slowest'' DNS exfiltration malware.

\section{Related Work} \label{sec:related-works}
To the best of our knowledge, research conducted thus far has focused on detecting DNS tunneling tools and largely overlooked low throughput malware communication. The prior research that mentioned the case of low throughput exfiltration malware failed to evaluate the approaches proposed on such cases. In this section we provide a brief review of related prior work which is summarized in Table~\ref{tbl:related-works}. It is important to note that none of the previous works evaluated their methods on malware samples or simulations, and only a few considered per domain classification in order to provide the immediate prevention of further leakage following detection.

\begin{table}[htb!]
	\centering
	\caption{Related Work Summary}
	\begin{tabular}{@{}lllll@{}}
		\toprule
		Study                                                & Year & Sample$^a$     & Evaluation$^b$                   & Technique$^c$               \\ 
		\midrule
		Paxson et al.~\cite{paxson2013practical}         & 2013 & PD & I      & RB           \\ 
		Kara et al.~\cite{kara2014detection}         & 2014 & PD & I      & RB           \\ 
		Sheridan et al.~\cite{sheridan2015detection}  & 2015 & PD & I      & RB     \\
		Wang et al.~\cite{wang2016combating}          & 2016 & PD & -       & UL \\
		Buczak et al.~\cite{buczak2016detection}      & 2016 & PU   & I+       & SL \\
		McCarthy et al.~\cite{mc2016data}             & 2016 & PD & I & DT \\
		Homem et al.~\cite{homem2016entropy}          & 2016 & PT   & I+       & UL \\
		Cambiaso et al.~\cite{cambiaso2016feature}    & 2016 & PT   & I       & UL\\
		Engelstad et al.~\cite{engelstad2017detection} & 2017 & PU   & Mob        & UL\\
		\bottomrule
	\end{tabular}
	\label{tbl:related-works} \par
	\footnotesize{$^a$ PD - per domain, PU - per user, PT - per time (series of packets), \\
		$^b$ I - Iodine only, I+ - At least two tunneling tools, Mob - Mobile tunneling \\
		$^c$ RB - Rule-based, UL - Unsupervised learning, SL - Supervised Learning, DT - Decision Theory }
\end{table}

Initial research in the field focused on specific DNS tunneling tools, mainly Iodine. Kara et al.~\cite{kara2014detection} analyze the difference in the distribution of TXT resource records requests between popular domains and domains used for tunneling with Iodine. Later versions of Iodine leverage other types of rich resource records (e.g., SRV, NULL). Sheridan et al.~\cite{sheridan2015detection} also focus on Iodine and rely on the fact that normal DNS usage will be followed by TCP communication to the resulting domain IPs. They analyze three scenarios: normal (no attack), passive Iodine, and active Iodine. They show a clear distinction in the amount of DNS requests versus TCP activity between the two scenarios.

The essence of a communication channel is in transferring data, as noted by Paxson et al.~\cite{paxson2013practical}, and in light of this, their scheme is based on collecting the total information sent per primary domain in terms of characters, query types, and time difference between requests. This information is then compressed, and if the result is greater than a certain threshold, the domain is flagged.

More recent work has focused on less specific tunneling tools using supervised learning. For example, Buczak et al.~\cite{buczak2016detection} also detect tunneling, although not only Iodine. The proposed method uses two different types of PCAP files: with and without tunneling; the tunneling is performed by one of three tools. A supervised random forest classifier is used to differentiate between the two classes. While Buczak et al. focused on per user detection, Homem et al.~\cite{homem2016entropy} focused on per time (series of packets); in their approach the changes in mean Shannon entropy between tunneling and benign traffic are learned. Although these approaches mention the existence of DNS exfiltration malware, they are only trained on high throughput DNS tunneling which may result in overfitting and ineffectiveness against a malware attack.

Based on these limitations, more recent research migrated away from supervised learning approaches to unsupervised learning in order to generalize the problem. Cambiaso et al.~\cite{cambiaso2016feature} use dimensionality reduction and mutual information to tackle the problem. In their work, higher order statistical features of the request and response sizes and the difference between consecutive requests are transformed to two principle components. Based on these components, a mutual information measure is computed to detect a ``weak independence'' relation between the original higher order statistics which is associated with tunneling. Other research performed by Engelstad et al.~\cite{engelstad2017detection} evaluates two anomaly detection techniques to detect tunneling over mobile DNS. In their work, a feature vector corresponds to a user PCAP and their findings show that OCSVM (one-class SVM) outperforms k-means in this setting. While offering a generic detection approach for one-class anomaly detection, only two techniques were considered (clustering and distance). Both of these studies fall short in two respects: (1) low throughput exfiltration of data that does not go over TCP is unlikely to create a major change (or any change at all) in consecutive samples and (2) focusing on per time / per user detection requires manual intervention to stop the data leakage. Thus, the detection rate of these solutions might be too low for large-scale DNS traffic.

A decision theory based approach for the problem was presented by McCarthy et al.~\cite{mc2016data}. Their work proposes a Markov decision process to infer the network node sensors to be activated for the detection of DNS exfiltration. This is done interactively, using a partially observable Markov decision process (POMDP) on the network graph, in which nodes correspond to hosts and edges correspond to channels. While this model is theoretically able to cover all possible attacks, it requires the ability to set sensors within the network which is not always possible (e.g., if the security system is a cloud solution service rather than an on-premise appliance within the network). Also, although the authors mention the existence of malware such as the FrameworkPOS, the approach is evaluated using the DETER testbed on the Iodine DNS tunneling tool alone, and hence its effectiveness was not proven on that malware.

Finally, we mention a proposed solution based on a new standard in which the difference between legitimate and malicious DNS tunneling is emphasized~\cite{wang2016combating}. In this work, Wang points out that the detection of tunneling is easy enough, since the volume of requests is high, however the distinction between types of tunneling is more difficult. He suggests a whitelist in the form of a legitimate tunneling directory in which reputable DNS tunneling services must register in advance; unregistered DNS services would be denied. This solution, if implemented, would decrease false positive rates for all other solutions. Although it focuses on the establishment of a post-detection whitelist, it does not specify a new detection system or discuss malware.

\section{Method} \label{sec:method}
\subsection{Overview}
The proposed method presents a constantly running process aimed at detecting and blocking  domains that are queried in streaming DNS traffic and used for malicious DNS data exfiltration. First, the streaming DNS traffic is collected and transformed to feature vectors corresponding to domains. Afterwards, a pre-trained one-class classifier is used to detect domains that exchange data over the DNS. Immediately after, requests to domains that are classified as used for data exchange are blocked indefinitely. In the rest of this section we describe the method in detail, starting with our basic assumptions (see subsection~\ref{subsec:assumptions}) and setting parameters (see subsection~\ref{subsec:parameters}), to each of the method's phases (see subsections~\ref{subsec:data-collection},\ref{subsec:feature-extraction}, \ref{subsec:anomaly-detection}).

\begin{figure} [htb!]
	\centering
	\includegraphics[width=\columnwidth]{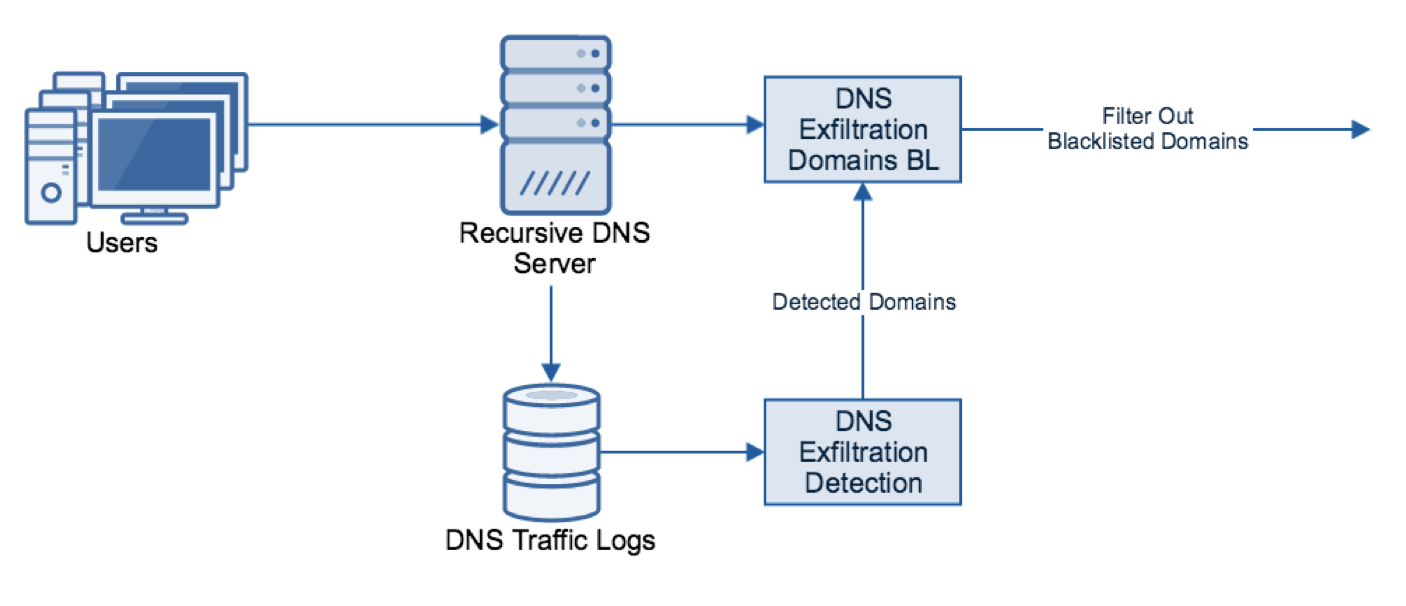}
	\caption{The high level design of a system in which a DNS server outputs its logs to be grouped as domains, classified as either malicious or benign, and blocked if classified as malicious. These are the only two requirements of the DNS server for employing the proposed detection system are DNS traffic logging and domain blacklisting capabilities.}
	\label{fig:deployment}
\end{figure}

Our method attempts to be as straightforward as possible to allow its implementation in DNS servers that are not necessarily designed to perform detection, as long they support logging the DNS traffic and domain blacklisting (as portrayed in Figure~\ref{fig:deployment}). 

In addition, the method relies on the capability to deploy the previously trained model regardless of the environment. This capability is supportedby selecting only statistical features (rates, averages, etc.) for the anomaly detection model, thereby avoiding non-windowed features such as the number of requests which may behave differently on different sized DNS servers. Moreover, the iForest model is serializable as a collection of binary trees and the anomaly score threshold.

\subsection{Assumptions} \label{subsec:assumptions}
Given the background presented in section~\ref{sec:background}, the method is based on the following assumptions:
\subsubsection{DNS Data Exchange Abnormality} \label{assum:abnormality} based on the general scheme for data exfiltration (subsection~\ref{subsec:scheme}), we establish the abnormality of DNS traffic when used for data exchange. We assume that domains that are used for data exchange over the DNS will likely be characterized with longer than average requests and responses, encoded payloads (subqueries), and a plethora of unique requests.
\subsubsection{Use of a Single Domain} based on recently discovered malware, we assume that a single domain is used for exfiltration, thus detecting its querying abnormality and denying its DNS requests are sufficient for stopping the leakage. A more complex case of multiple domains is discussed in the future work section (section~\ref{sec:discussion}).

\subsection{Parameters} \label{subsec:parameters}
The proposed method has three tunable parameters (summarized in Table~\ref{tbl:parameters}).
\begin{table}[htb!]
	\centering
	\caption{Proposed Method's Parameters}
	\begin{tabular}{@{}lll@{}}
		\toprule
		Name & Description & Example Value     \\
		\midrule
		$\nu$  & Acceptable rate of false positive domains & $2\cdot10^{-5}$ \\
		$\lambda$  & \begin{tabular}{@{}c@{}}Frequency of data collection \\ and classification [minutes]\end{tabular}  & 15                \\
		$n_s$ & \begin{tabular}{@{}c@{}}Window size for inspection \\ (i.e., number of data collection sets)\end{tabular} & 24 \\
		\bottomrule
	\end{tabular}
	\label{tbl:parameters}
\end{table}

The settings of the parameters may influence the detection rate and its latency. Based on our evaluation, we recommend the following:
\begin{enumerate}
	\item \emph{Setting $\nu$:} A higher acceptable rate of false positives (denoted as $\nu$) will potentially increase the detection rate at the cost of additional false positive domains. We recommend setting it to a value not higher than 0.1\% (due to threat scarcity) and decreasing it to the point when the acceptable rate of false positive domains to be reviewed by a security expert is reached. In our false positives analysis (see Section~\ref{sec:false_positives}), setting $\nu=2\cdot10^{-5}$ yielded only 18 false positives domains on a large-scale DNS traffic with a peaking rate of 47 million hourly queries over the course of six days. Moreover, since the number of false positives domains decreased exponentially since the start execution, we expect that given a white-listing of former false positives, the rate would effectively be much lower.
	\item \emph{Setting $\lambda$:} A lower frequency of data collection and classification (denoted as $\lambda$) will allow better detection latency, e.g., up to $\lambda$ minutes since exfiltration starts. However, it will require that the feature extraction be applied on $n_s \cdot \lambda$ minutes of logs each $\lambda$ minutes. We therefore, recommend setting it to as low as the system's storage, processing power, and memory allocation.
	\item \emph{Setting $n_s$:} A longer window size of inspection in units of $\lambda$ minutes (denoted as $n_s$) allows detecting "low and slow" attacks, even if they take place over several hours (e.g., ten credit card numbers are extracted over a period of six hours). In the same way as $\lambda$, it should be increased as much as possible within the limitations of storage and memory.
\end{enumerate}

Moreover, the trade-off between immediate blocking and detecting a silent attack should be reflected in setting both $n_s$ and $\lambda$. For example, a hotel's secured wireless network that wishes to focus on blocking unpaid Wi-Fi access based on DNS tunneling may choose $\lambda=1,\,n_s=10$, while a point of sale network may use $\lambda=60,\,n_s=10$.

\subsection{Data Collection} \label{subsec:data-collection}
The main goal of the data collection phase is to generate a compact representation of per domain traffic that will facilitate a subsequent efficient feature extraction process.

This phase starts by processing the streaming DNS traffic every $\lambda$ minutes and represents each DNS log line as the following tuple:
\[
\langle Q,R,T\rangle_{j}\ ,
\]
such that $Q$ contains the full query name (e.g., google.com.), $R$ contains the full list of response record values (e.g., 8.8.8.8), $T$ contains the queried record type (e.g., "A"), and $j$ is the log line index. Specifically for the case of non-existent responses (a.k.a. NXDOMAIN status), $R$ is assigned with an empty string. Based on this representation we group DNS logs by primary domains.

A primary domain name is hereby defined as the concatenation of the second level domain~\footnote{https://en.wikipedia.org/wiki/Second-level_domain}, and top level domain~\footnote{https://en.wikipedia.org/wiki/Top-level_domain}. For example, for the fully qualified domain name login.example.org.de. the second level domain is example. and the top level domain is org.de.; hence the primary domain is example.org.de.. The main use of a primary domain name is for domain registration. Therefore, by using it as a sample point for a classification task it assigns a class for each registration, and is effective against domains that were registered merely for malicious acts. The primary domain can be extracted from each log line (as explained above) using a function, $prim$, as follows:
\[
prim\left(\langle Q,R,T\rangle_j\right) = P_j
\]
where $P_j$ is the primary domain for the j-th log line. Using the $prim$ function, log lines can then be grouped by their primary domain, as well as by their discrete time frame of collection $t$:
\[
L^{P_i}_{t} = \left\{ \langle Q,R,T\rangle_j\, | \,prim\left(\langle Q,R,T\rangle_j\right) = P_i \right\} \ .
\]
This grouping of DNS logs allows for efficient scanning of consecutive logs, as each log can be collected once but used $n_s$ times using a sliding window. For example, a log line that will be collected at time $t_i$ will be consumed at times $[t_i,..., t_i + n_s)$.  This sliding window over the recent $n_s \cdot \lambda$ minutes proves to be useful in the feature extraction phase which follows.

\subsection{Feature Extraction} \label{subsec:feature-extraction}
The feature extraction phase works once every $\lambda$ minutes on a per domain sliding window, i.e. a combination of the last $n_s$ collected logs:
\[
W^{P_i}_{t_{now}} = \left\{ L^{P_i}_{t} \, | \, t_{now} - n_s <= t < t_{now}  \right\}
\]

We define a feature extraction function, $fe$, to transform a sliding window, $W^{P_i}$, to a feature vector that represents a specific primary domain in a certain sliding window:
\begin{equation}\label{eq:feature-vector-function}
fe\left(W^{P_i}_{t_{now}}\right) = \langle P_i, Ent, NI, Uniq, Vol, Len, LMW\rangle \ ,
\end{equation}
where each of the following features is a computed function on the domain $P_i$ in the time frame $t_{now}$:
\begin{enumerate}
	\item $E$ corresponds to the character entropy (see~\ref{subsec:shannon}),
	\item $NI$ corresponds to the non-IP type ratio (see~\ref{sec:nonip_ratio}),
	\item $Uniq$ corresponds to the unique query ratio (see ~\ref{sec:unique-query-ratio}),
	\item $Vol$ corresponds to the query volume (see~\ref{sec:query_volume}),
	\item $Len$ corresponds to the query average length (see ~\ref{sec:query_length_avg},
	\item and $LMW$ corresponds to the ratio between the length of the longest meaningful word and the subdomain length (see~\ref{sec:lmw}).
\end{enumerate}
We now elaborate on these features and how to compute each of them.

\subsubsection{Character Entropy} \label{subsec:shannon}
The notion of entropy, or information density, can be interpreted as measuring the average uncertainty of a letter $A_{n}$ given $\{A_{1},..,A_{n-1}\}$ ~\cite{shannon1951prediction}. Generally, entropy is computed over a discrete random variable $X$ using the formula:
\[
H(X) = - \sum_{i=1}^{n} \Pr(x_i) \cdot \log \Pr(x_i)
\]
where $\Pr(x_i)$ is the probability of the i-th symbol of information (e.g., a character) in the series $X$ composed of up to $n$ symbols. Among other uses, entropy is widely used as a heuristic for the detection of encryption in a stream of bits~\cite{lyda2007using, rossow2013provex, dorfinger2010real}.
We compute the entropy on per domain queries to detect encrypted or encoded querying behavior. Formally, if the random variable $X$ is a series of characters, and the allowed symbols are the letters, digits and hyphens (LDH) as in the Request for Comments (RFC) of the DNS protocol~\cite{rfc1034}, then:
\[
H(X) = - \sum_{x_i \in \text{LDH}} \Pr(x_i) \cdot \log \Pr(x_i)
\]
and the feature is computed as follows:
\[
E(W^{P_i}_{t_{now}}) = H(Q_1 || .. || Q_{m})
\]
where $Q_1 || .. || Q_{m}$ is a concatenation of the fully qualified domain names in the log lines of  $W^{P_i}_{t_{now}}$.

\subsubsection{IP-Hostname Exchange RR Type Distribution} \label{sec:nonip_ratio} According to~\cite{herrmann2013behavior}, 99.4\% of the requested resource records (RRs) are of the following types: A~(IPv4), AAAA~(IPv6), and PTR (reverse lookup pointers).
Given the main function of DNS as the "phonebook of the Internet"\footnote{https://en.wikipedia.org/wiki/Domain_Name_System}, it would make sense to assume that RR types that relate domains to IP addresses represent the vast majority of DNS lookups. However, these RRs are restricted to a short response length (i.e., up to the length of an IP address) compared to other RR types (e.g., TXT, SRV), and thus the distribution of RR types may be different in a domain used for data exchange. The feature we create computes the rate of A and AAAA records for per domain in time, i.e.,
\[
NI(W^{P_i}_{t_{now}}) = \frac{\sum_{W^{P_i}_{t_{now}} \mid (T="A" \wedge T="AAAA")}  1} {\sum_{W^{P_i}_{t_{now}}} 1}\ .
\]
\subsubsection{Unique Query Ratio} \label{sec:unique-query-ratio} A domain whose subdomains are used as messages is not likely to repeat them. Therefore, when comparing domains used for exfiltration to normal domains we expect to see a much higher unique query ratio for the latter. The feature is computed as follows:
\[
Uniq(W^{P_i}_{t_{now}}) = \frac{|\{ Q | Q \in W^{P_i}_{t_{now}}\}|} {\sum_{W^{P_i}_{t_{now}}} 1}\ .
\]
\subsubsection{Unique Query Volume} \label{sec:query_volume} In a normal setting, DNS traffic is rather sparse as responses are largely cached within the stub resolver. However, in the case of data exchange over the DNS, the domain-specific traffic is expected to avoid cache by non-repeating messages, or short time-to-live in order for the data to make it to the attacker's server. Avoiding cache as well a lengthy data exchange, might result in a higher volume of requests compared to a normal setting. The feature is computed as follows:
\[
Vol(W^{P_i}_{t_{now}}) = |\{ Q | Q \in W^{P_i}_{t_{now}}\}|\ .
\]
\subsubsection{Query Length Average} \label{sec:query_length_avg} As a complementary to the volume feature and given a query size limitation, there is a trade-off between the volume of queries and their length, hence making it an effective feature for detection.
\subsubsection{Longest Meaningful Word Over Domain Length Average} \label{sec:lmw} For each primary domain, each subdomain is decomposed to its hierarchy labels ordered by their length. Starting from the longest to the shortest substring, an English dictionary word lookup is performed over the current substring. If the lookup succeeds (i.e., the substring is a valid English word) - the length of the substring is taken as the \emph{longest meaningful word} (LMW) length. This length is then divided by the length of the subdomain and averaged out over all of the subdomains. While English is not the only language used on the Internet, this may help distinguish domains with a large number of to domains to either readable, or non-readable.

\subsection{DNS Data Exchange Detection} \label{subsec:anomaly-detection}
The detection of DNS data exchange for an input feature vector is computed using an anomaly detection model, namely, Isolation Forest~\cite{liu2008isolation}. The Isolation Forest model is a one-class classifier (i.e. trained only on existing legitimate data) and can detect anomalous behavior on future data. Therefore, two aspects should be discussed here: (1) training the model and (2) applying the model on new data.

The training phase takes a set of previously collected feature vectors and outputs an anomaly model, which is essentially a function that acts on a sample and outputs an anomaly score. In our case, the input to the model is the set $W^{P_i}_{t}$ for every domain $i$ and for $t$ at a specific time period (e.g., the previous day) and the output is the anomaly score ranging from 0 to 1. The anomaly score is a function of the contamination rate, denoted as $\nu$, which refers to the fraction of noise in the data. The output of the training phase is the anomaly model together with $T_s$, the anomaly score threshold, which will be applied on new data.

As a new sample arrives, the model is applied to it such that each sample is assigned with a score, $s$, using a function $iforest$, as follows:
\begin{equation}
iforest\left(fe\left(W^{P_i}_{t_{now}}\right)\right) = s \ ,
\end{equation}
where $fe$ is the feature extraction function from Eq.~\ref{eq:feature-vector-function} that translates a sample to a feature vector. If the score exceeds the anomaly score (i.e., $s > T_s$), the sample is considered anomalous and the domain it refers to will be marked as a domain used for data exchange over the DNS.

\subsection{Blocking Malicious Domains} \label{subsec:blocking-phase}
As mentioned before, domains that are anomalous in terms of traffic (i.e., used for data exchange) can be divided into two categories: malicious and legitimate. Generally, when the system is deployed in a network it will readily find the legitimate (and possibly also illegitimate) services that use DNS for data exchange. Once these domains are mapped and removed by a security expert, any new domain that comes up in the anomaly detection phase is considered malicious data exfiltration and blocked immediately.

\section{Evaluation} \label{sec:evaluation}
The evaluation focuses on two primary goals: the detection of low throughput malware exfiltration, and the detection of high throughput DNS tunneling. To that end, we introduce our dataset of DNS traffic (see subsection~\ref{subsec:datasets}) comprised of benign traffic as well as DNS exfiltration test subjects (see subsection~\ref{sec:test_subjects}). The test subjects include: (1) Iodine and (2) Dns2tcp (as high-throughput tunneling tools), and (3) FrameworkPOS and (4) Backdoor.Win32.Denis (as DNS exfiltration malware). 

The detection and false positive rates for the proposed method are reviewed for each of the test subjects when placing a strong limit on the acceptable false positive rate parameter (i.e., less than $2 * 10^{-5}$). Based on a manual use-case classification conducted on the false positive detection (see Table~	\ref{tbl:fps}), we conclude that the majority of false positive domains fall under the use-case of legitimate data exchange over the DNS (e.g., Antivirus signature lookup) and should therefore be white-listed by a security expert upon detection in order to prevent future false alarms. Therefore, given that former are white-listed, we analyze the false positive rates over time (instead of per domain) by measuring the number of new falsely detected domains per day (see section~\ref{sec:false_positives}). 

Furthermore, the proposed method is compared with re-implementations of two recently published methods~\cite{cambiaso2016feature,homem2016entropy}, that are designed for the detection of tunneling in order to measure the effectiveness on our test subjects that include of both tunneling and malware. This comparison is the basis for suggesting that avoiding a malware evaluation in previous works is not an additional overlooked test, but rather an undealt case. 

\subsection{Dataset} \label{subsec:datasets}
Our main dataset (denoted as $DS$) is a single week of DNS traffic collected from a subset of recursive DNS resolvers operated by Akamai Technologies. $DS$ is considered as a large-scale DNS traffic sample with at least a hundred thousand end-users, 35 million hourly requests on average with a standard deviation of 74 million hourly requests (further statistics appear in Table~\ref{tbl:datasets}).

Arguably, this makes $DS$ an ideal candidate for the evaluation of our proposed method on large-scale DNS traffic. In contrast, both of the recent works that we use for comparison were originally designed and evaluated on small-scale DNS traffic. The evaluation of both works on the large-scale dataset, $DS$, yielded sub-par results that were significantly different from their reported results, even for the DNS tunneling class. Therefore, to conduct an additional fair test, we introduce $DS_{partial}$. $DS_{partial}$ contains the traffic of a subset of $DS$ users' traffic. Therefore, it is slightly smaller than one tenth of $DS$, thus making it more suitable for comparison with prior research.

% \usepackage{booktabs}
\begin{table}[htb!]
	\centering
	\caption{Datasets Description of Hourly Queries}
	\begin{tabular}{@{}llllll@{}}
		\toprule
		& mean           & std            & min            & median         & max            \\ \midrule
		$DS$           & $3.5 \cdot 10^{7}$ & $7.8 \cdot 10^{6}$ & $2.2 \cdot 10^{7}$ & $3.7 \cdot 10^{7}$ & $4.7 \cdot 10^{7}$ \\
		${DS}_{partial}$ & $2.9 \cdot 10^{6}$ & $4.9 \cdot 10^{5}$ & $1.8 \cdot 10^{6}$ & $2.7 \cdot 10^{6}$ & $4.0 \cdot 10^{6}$ \\ \bottomrule
	\end{tabular}
	\label{tbl:datasets}
\end{table}

During the single week of DNS traffic of which $DS$ and $DS_{partial}$ are composed, we have injected the traffic of real malware communication and DNS tunneling tools which will be used to test model but not to train it (this will be fully explained in subsection~\ref{sec:test_subjects}). Due to the scarcity of data exchange over the DNS, we can assume that aside from our injected traffic there are no additional DNS data exchange cases on either dataset.

\begin{figure} [ht]
	\centering
	\includegraphics[width=\columnwidth]{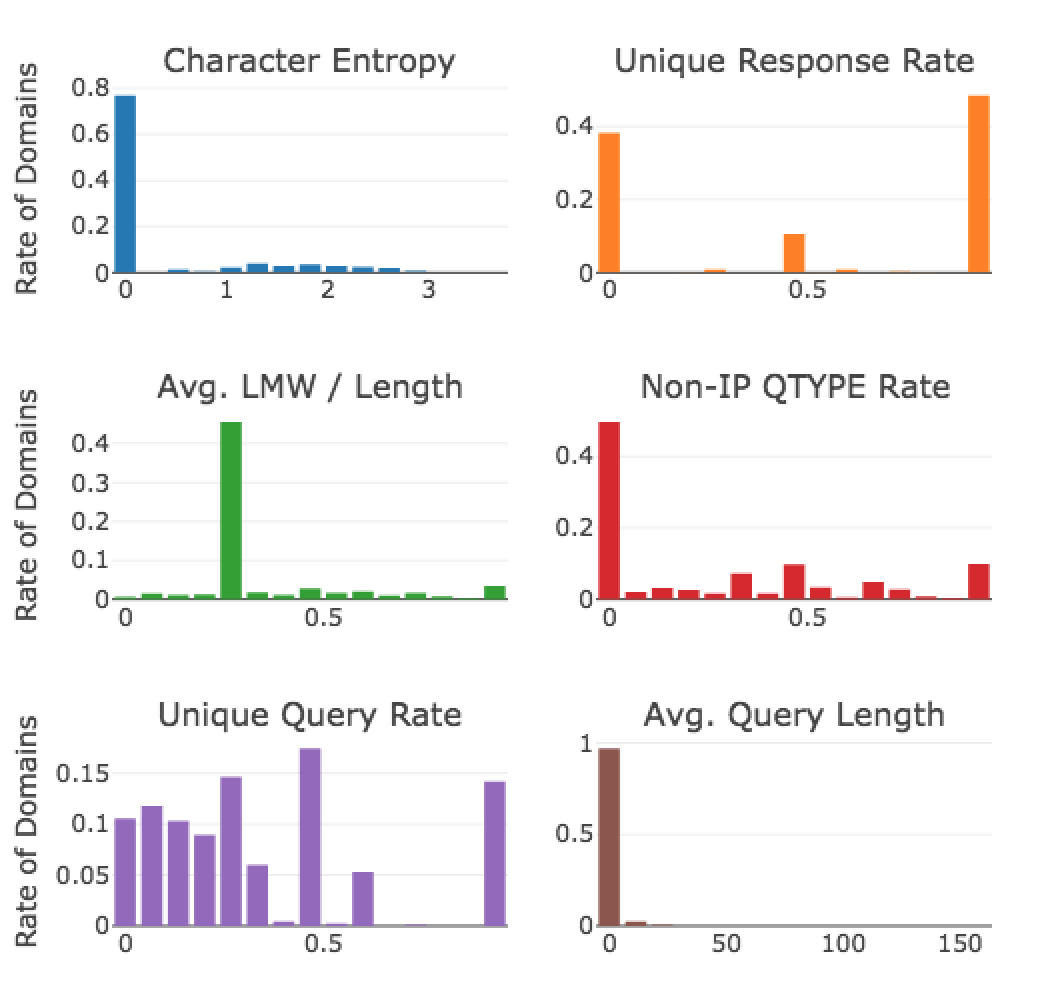}
	\caption{The probability density function (PDF) for each of the extracted features on the first day included in the dataset (presumed to contain no attacks).}
	\label{fig:features_pdf}
\end{figure}

Each of the domains within the $DS$ dataset is represented as a feature vector, as displayed in subsection~\ref{subsec:feature-extraction}. The first day included in $DS$, which is presumed to contain only clean traffic, is explored in Figure \ref{fig:features_pdf}. As discussed in subsection~\ref{assum:abnormality}, we expect domains that are used for data exfiltration to have corresponding feature vectors that are appear as outliers with regards to normal domains' feature vectors. For example: Iodine DNS tunneling has a mean query length that is larger than 100 characters whereas less than 0.01 of normal domains behave in this manner. Also, with Iodine's default encoding set to Base128, the character entropy normally exceeds 3.0 whereas in the case of clean traffic less than 0.05 of the traffic follows this behavior.

\subsection{Test Subjects}\label{sec:test_subjects}
There are four test subjects which will be tested against compared previous works and the proposed method. All of them are direct simulations of DNS tunneling software and DNS exfiltration malware as executed ``in the wild.'' Although this relatively small number is not ideal for the evaluating the detection rate (e.g., using a receiver operating characteristic curve), it is the most unbiased way of testing these methods.  

\subsubsection{FrameworkPOS} \label{sec:frameworkpos}
The FrameworkPOS malware~\cite{rascagneres2016new} was used in a targeted attack on the American retailer, Home Depot, back in 2014. Over the course of six months, the malware leaked information associated with not less than 56,000,000 credit card account numbers. The malware operates autonomously such that it extracts newly available credit card information from the compromised machine's memory, encodes the data and sends it to a remote server. The data is sent using a DNS query to a dedicated domain of the form: $\langle encoded\_credit\_card\rangle.domain.com$. Although FrameworkPOS' \cnc server was shut down after the Home Depot campaign, the details mentioned above allow us to construct an accurate simulation.

Our simulation of FrameworkPOS consists of a dedicated domain (simply referred to as frameworkpos.com.) Using this domain, we output DNS queries of the form: $\langle encoded\_credit\_card\rangle.frameworkpos.com$ at an average rate of three queries per second (proportional to 56 million credit card account numbers in six months). Also, the queries' resource record type is set to ``A'' (IPv4 request) as was done in the Home Depot campaign.

\subsubsection{Backdoor.Win32.Denis} 
The Trojan malware Backdoor.Win32.Denis~\cite{shulmin2017tunn4comm} (referred to as Denis in this paper) was discovered by Kaspersky Labs in 2017. Before its discovery, Denis was used in Operation Cobalt Kitty, a large-scale APT in Asia~\cite{dahan2017cobaltkitty}. Throughout its execution, it enables an intruder to manipulate the file system, run arbitrary commands and run loadable modules. In contrast to FrameworkPOS (see subsection~\ref{sec:frameworkpos}) which uses the DNS as a payload delivery channel, Denis uses the DNS as a bidirectional communication channel with its \cnc server. It has 16 predefined instructions to allow the \cnc operator to take control of a compromised machine.

Among these instructions is a ``keep-alive'' instruction that is sent routinely throughout Denis' execution. This keep-alive instruction is manifested as a query with a ``NULL'' resource record type approximately every 1.5 seconds, and it contains the ID of the compromised machine and a unique identifier to reach the \cnc server regardless of DNS caching.

Based on the above, the simulation of Denis consists of a dedicated domain (z.teriava.com) and DNS queries of the form: \[\resizebox{.95\hsize}{!}{vL0VugAAAAAAAAAAAAAAAAAAAAAAAAvV5.z.teriava.com}\] where the last three characters of the subdomain act as an anti-caching mechanism. The queried resource record type is NULL (i.e., to allow a longer answer than an IPv4 address) and the requests are sent every 1.5 seconds exactly as took place in the Cobalt Kitty operation's security report analysis.

\subsubsection{Iodine}
Iodine is an open-source DNS tunneling tool. It is commonly used to bypass Wi-Fi payment by browsing the Internet over the DNS. Our use, for evaluation purposes, is a single hour of Web browsing through a SOCKS proxy that is tunneled over the DNS using Iodine.

\subsubsection{Dns2Tcp}
Similarly to Iodine, Dns2tcp is also an open-source tunneling tool. It is commonly used to tunnel general purpose TCP sessions. The Dns2tcp use case in our evaluation was a download of a 30-kilobyte file from an FTP server.

\subsection{Detection Rate} \label{sec:detection_rate}
The model is trained in an unsupervised learning manner, and is never exposed to attacks during its training. Aside from the comfort of not having to set labels for the training process and the analysis, a successful detection rate for both tunneling and malware will imply the model well generalizes the problem of detecting both classes.

The training phase of our method is applied to the first day of the $DS$ dataset, which only contains benign traffic. For the sake of efficiency, instead of training over $24 \cdot 3.5 \cdot 10^7 \approx 8.4 \cdot 10^8$ records, we discard primary domains with less than ten sub-domains in the last $\lambda \cdot n_s$ minutes from both the training and the execution phases. Although this filter ignores exfiltration at a rate lesser than 2.5kb per $n_s \cdot \lambda$ (at most 10 queries of 255 bytes each), we manage to reduce our training set to $2.1 \cdot 10^{5}$, i.e., 0.25\% of its original size. 

The time window model parameters are set to $n_s=6, \lambda=60$, thus the method remains capable of detecting attacks that are as slow as $0.11b/s$ and under a worse case analysis will block and report their activity up to six hours after their first collected data. Effectively however, for two real-life malware that are evaluated in section~\ref{sec:evaluation} they are caught immediately within the first $\lambda$ minutes following their execution. 

The acceptable false positive rate is set to $\nu=2 \cdot 10^{-5}$, i.e., it will only detect domains that are at least as anomalous as the four most anomalous domains in the benign traffic. The training phase with the above parameters yields an anomaly score threshold of $T_s=0.653$. Based on the output model and $T_s$, we apply the model to new traffic in $DS$ that was not used for training.

The execution phase of the model works likewise. Every $\lambda=60$ minutes, domains that had at least ten subdomains within the last $\lambda * n_s = 360$ minutes become candidates for classification. The candidate domains' logs are transformed to feature vectors as described in Section~\ref{sec:method} and the model assigns them an anomaly score. Based on the training phase, domains whose anomaly score exceeds $T_0.653$ will be regarded as domains that are used for exfiltration and will be immediately blocked. 

\begin{figure} [ht]
	\centering
	\includegraphics[width=\columnwidth]{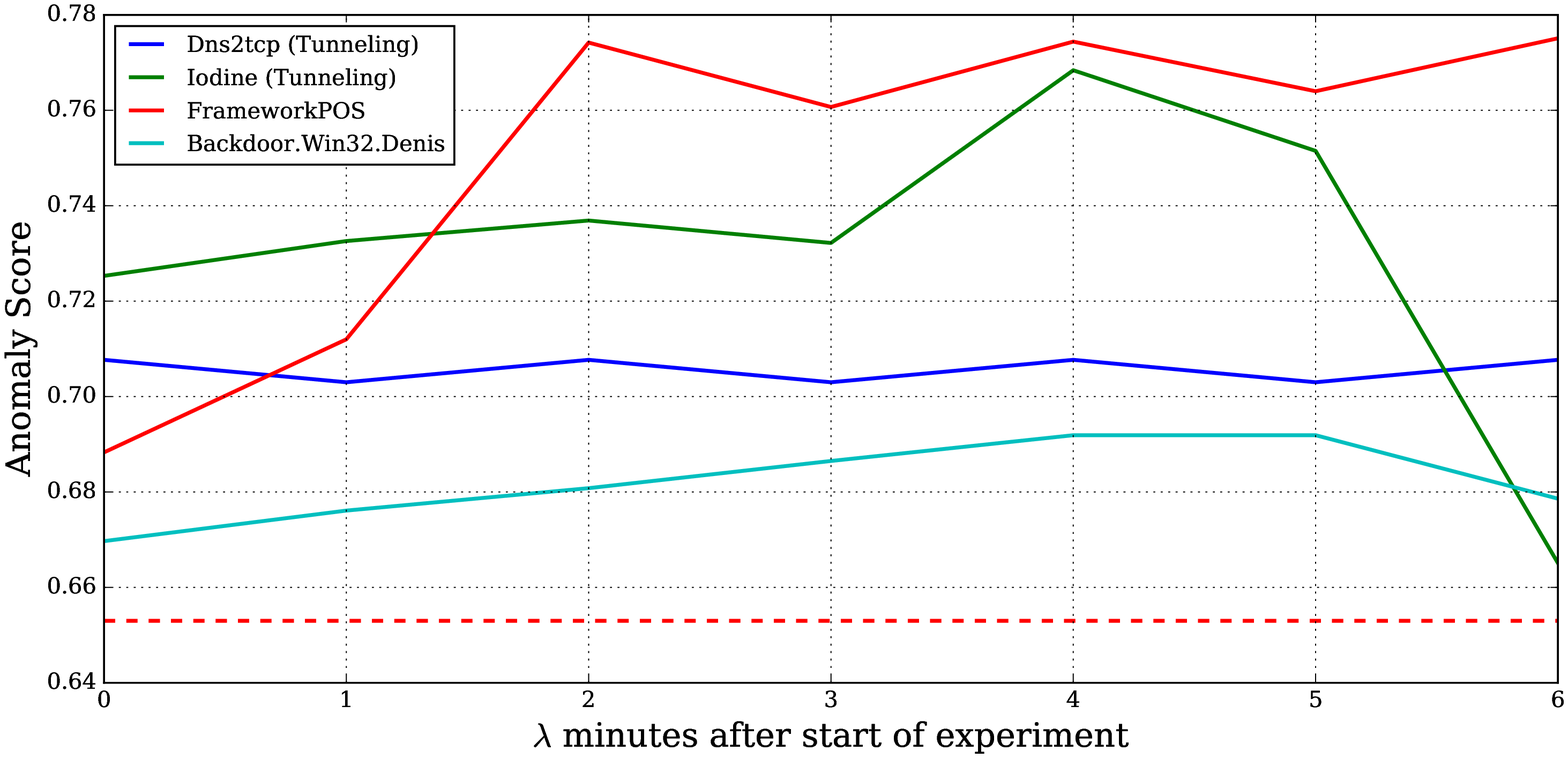}
	\caption{The anomaly score of the domains corresponding to Dns2tcp, Iodine, FrameworkPOS and Backdoor.Win32.Denis over time. The horizontal dashed line stands for anomaly score threshold (i.e., $T_s = 0.653$) when trained with an acceptable rate of false positives set to $\nu=2 \cdot 10^{-5}$, $n_s=6, \lambda=60$.}
	\label{fig:anomaly_score}
\end{figure}

The first part of our method evaluation focuses on the rate of successfully detected test subjects. The anomaly score for each of the test subjects is sampled every $\lambda$ minutes as of the traffic injection start. The expected result is for every test subject's anomaly score to exceed the given $T_s=0.653$ threshold in at least a single sample in order to be regarded as a successful detection. The results presented in Figure~\ref{fig:anomaly_score} indicate that all of the test subjects that were successfully detected by our method during their execution.

\subsection{False Positives Rate} \label{sec:false_positives}
\begin{figure} [htb!]
	\centering
	\includegraphics[width=\columnwidth]{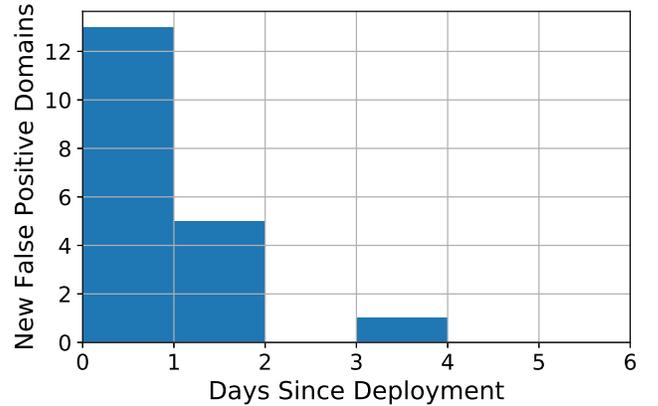}
	\caption{The number of new mis-classified domains (referred to as false positives) per day. The first day of traffic included in $DS$ to which the model is applied is regarded as zero.}
	\label{fig:false_positives}
\end{figure}

In addition to evaluating the detection rate (see subsection~\ref{sec:detection_rate}), we focus on evaluating the rate of false positives. Because our method classifies domains within a discrete time period (i.e., $\lambda \cdot n_s$ minutes), we attempt to evaluate it based on the overall number of domains that were mis-classified as malicious exfiltration. For that task, we use the $DS$ dataset. In the same manner used in the detection rate evaluation, we train on the traffic data from the first day included in $DS$ and apply our model to the rest of the traffic included in $DS$. The number of new false positive domains over time drops significantly and converges to at most a single domain per day only after only two days in which the model was used for classification (see Figure~\ref{fig:false_positives}).

The list of mis-classified domains is mainly comprised of legitimate security services that mis-use the DNS protocol for data exchange (see Table \ref{tbl:fps} for a summarized categorization of falsely detected domains). In our evalaution these domains are regarded as false positives, even though they are clearly used for data exchange, because they are assumed to be intentionally installed by the user and should therefore not be blocked. In a real-world implementation, one can assemble a whitelist of these domains before applying the automatic blocking, in order to avoid false shutdown of legitimate services.

% \usepackage{booktabs}
\begin{table}[htb!]
	\centering
	\caption{Classification of False Detections}
	\begin{tabular}{@{}lll@{}}
		\toprule
		Category                  & Frequency    & Examples                                                                                               \\ \midrule
		Security Services Lookups & 12 out of 18 & \begin{tabular}[c]{@{}l@{}}sophosxl.net, \\ l2.nessus.org,\\ avts.mcafee.com,\\ a.e.e5.sk\end{tabular} \\
		General Services Lookups  & 4 out of 18  & \begin{tabular}[c]{@{}l@{}}kr0.io,\\ dsipsl.net,\\ drtst.com,\\ dsipsl.net\end{tabular}                \\ 
		Others (*)                & 2 out of 18  & \begin{tabular}[c]{@{}l@{}}jacksonriverdev.com,\\  groupinfra.com\end{tabular}  \\ \midrule                       
	\end{tabular}
	\label{tbl:fps}\par
	\footnotesize{(*) primary domains with a large number of subdomains}
\end{table}

\subsection{Comparison with Prior Research}
After evaluating our method's detection rate (see subsection~\ref{sec:detection_rate}) and false positives rate (see subsection~\ref{sec:false_positives}), we compare our method to methods proposed in recently published works.

\subsubsection{Feature Transformation and Mutual Information for DNS Tunneling Analysis}
Cambiasso~et~al.~\cite{cambiaso2016feature} (denoted as C16) measures the mutual information (MI) of transformed features once in every 10,000 DNS queries. In the case of a sufficiently low MI, the model implies the existence of a DNS tunnel.

The comparison is based on an implementation of C16 over the $DS$ dataset, consisting of slightly more than 10,000 users. Apart from the MI threshold that was re-adjusted due to noise in our network, the other configurable parameters are set to $n_s=10000, r=500, \alpha=\frac{1}{1000}$ exactly as reported in C16. In addition, because C16 was initially designed for medium-scale DNS traffic we test it on the $DS_{partial}$ dataset, which includes all of the test subjects present on the $DS$ dataset but contains less than one tenth of its traffic.

\begin{figure} [ht]
	\centering
	\includegraphics[width=\columnwidth]{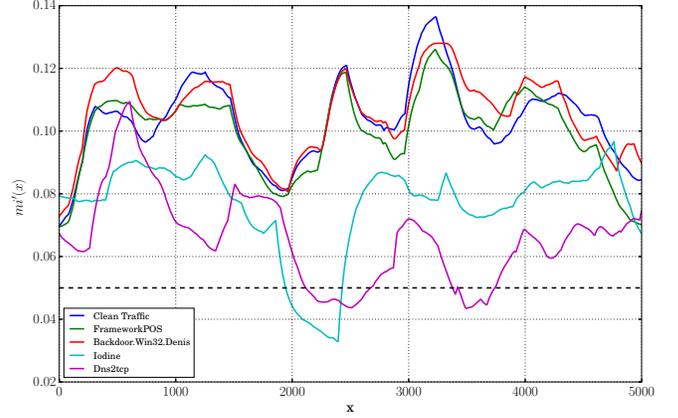}
	\label{fig:cambiasso}
	\caption{The smoothed mutual information measure (y-axis) of transformed features per feature index (x-axis). The feature index increments every 10,000 DNS queries and refers (in a sliding window) to the most recent 5M DNS queries. The plotted functions represents 5,000 feature indices (i.e., 50M DNS requests) of clean traffic, FrameworkPOS, Backdoor.Win32.Denis, Iodine and Dns2TCP. The horizontal dashed line (y=0.05) is the upper threshold for DNS tunneling detection.}
\end{figure}

The results of C16 with FrameworkPOS, Denis, and the DNS tunneling tools are presented in Figure~\ref{fig:cambiasso}). Each of the full-line plotted functions represents the MI over time, such that an MI below the dashed horizontal line represents DNS tunneling. While the results point to the clear detection of the high throughput tunneling cases (Iodine, Dns2tcp), the traffic of FrameworkPOS and Denis malware remains indistinguishable from clean traffic.

\subsubsection{Entropy-based Prediction of Network Protocols in the Forensic Analysis of DNS Tunnels}
Homem~et~al.~\cite{homem2016entropy} (denoted as H16) is designed to distinguish DNS tunneling from clean traffic based on the mean differences (referred to as ``MeanDiff'') between the normal average entropy (${\mu}_{X}$) and that of the latest N packets (${\mu}_{Y}$). The calculation of MeanDiff for each sample of N packets ($Y$) is computed as follows:
\begin{equation}
m(X,Y) = |{\mu}_X - {\mu}_Y| \ .
\end{equation}

A classifier is then trained on the single feature $m(X,Y)$ to determine if a new series of packets, $Y$, includes a tunnel. Because a concrete threshold, $t$, is not mentioned (e.g., $Y$ is a DNS tunnel if and only if $m(X,Y) > t$), we assume that $t$ is at least as large as the standard deviation of benign traffic entropy to avoid over-sensitivity of the model. For the sake of setting a threshold that will match our traffic, we calculate the standard deviation of query names entropy ($\sigma_{X}$) using a set of 50,000 query names without any tunneling:
\begin{equation}\label{eq:homem-std}
\sigma_{X} = 0.5369
\end{equation}
Based on that, we now assume that any series of packets will not be regarded as a tunnel if $m(X,Y) \leq \sigma_{X}$. Although we assume the classifier's threshold is more restrictive (i.e., significantly larger than $\sigma_{X}$), the above assumption assists us make a point about the possible detection of low throughput malware.

\begin{figure} [htb!]
	\centering
	\includegraphics[width=\columnwidth]{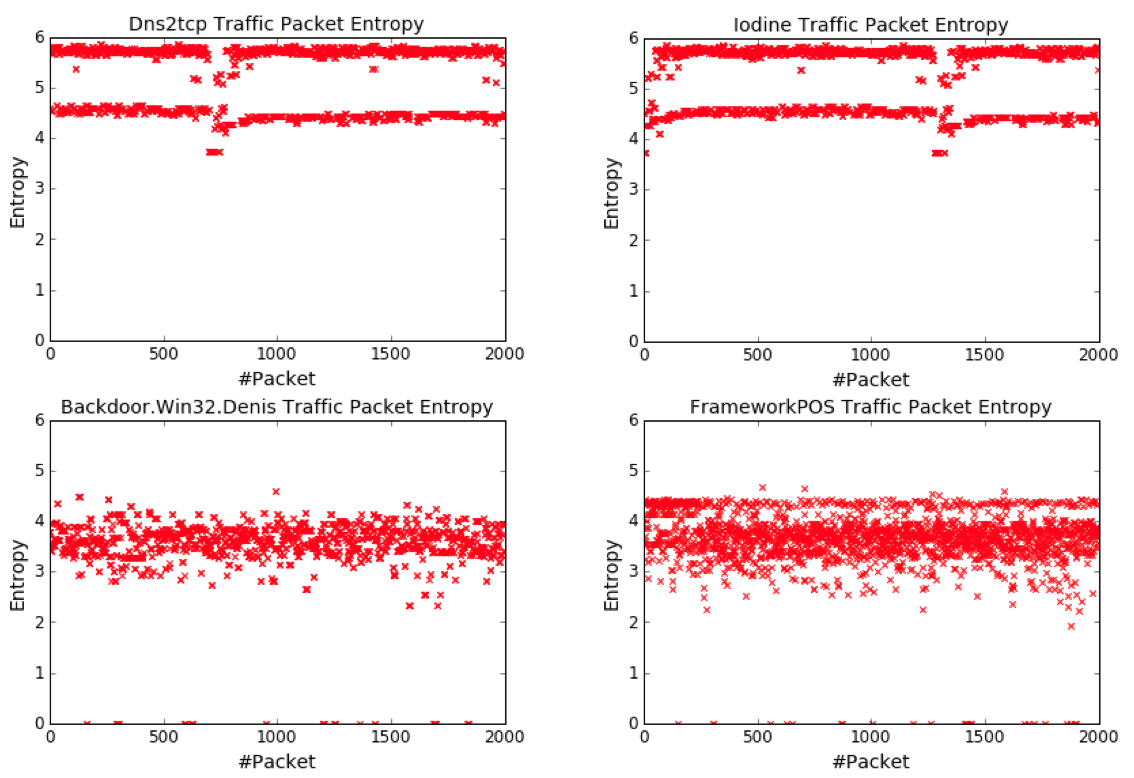}
	\caption{The query name entropy per query name over a series of 2,000 consecutive queries of clean traffic combined with Dns2tcp (upper left), Iodine (upper right), Backdoor.Win32.Denis (lower-left), and FrameworkPOS (lower-right).}
	\label{fig:homem}
\end{figure}

The evaluation of H16 was initially performed on twenty network captures of an unknown size. We test it on the $DS_{partial}$ dataset to allow it to become more sensitive (as the legitimate traffic proportion is decreased by $90\%$ compared to $DS$). The results on the test subjects in $DS_{partial}$ appear in Figure~\ref{fig:homem}. We evaluate $X$ based on 50,000 packets resulting in $\mu_{X}=3.558$ and 2,000 packets for each $Y$ test subject. 

The summarized results of H16 appear in Table~\ref{tbl:homem}. Recalling the standard deviation value (Eq.~\ref{eq:homem-std}), we see that while the DNS tunneling tools generate MeanDiff values over $3\sigma$, the values for the exfiltration malware are well within the standard deviation and are therefore undetectable using this method.

\begin{table}[htb!]
	\centering
	\caption{MeanDiff value of entropy based on H16}
	\begin{tabular}{@{}ll@{}}
		\toprule
		Detection Experiment & MeanDiff        \\
		\midrule
		Backdoor.Win32.Denis & 0.017118 \\
		FrameworkPOS         & 0.103525 \\
		Iodine               & 1.670904 \\
		Dns2tcp              & 1.684665 \\
		\bottomrule
	\end{tabular}
	\label{tbl:homem}
\end{table}

\subsection{Summary of the Results and Comparison with Prior Research}
To summarize, we used a single week of DNS traffic logs which were injected with the traffic of four test subjects, in order to compare our method of with two recently published methods, namely C16 and H16. The expected results was a successful detection of all subjects that simulated the following:
\begin{enumerate}
	\item Web browsing over the DNS using Iodine
	\item File transfer over the DNS using Dns2tcp
	\item Credit card account number theft using FrameworkPOS malware
	\item Trojan communication over DNS using Backdoor.Win32.Denis malware
\end{enumerate} 

A summary of the results (Table~\ref{tbl:results}) shows that our proposed method effectively detected all of the above scenarios. However, while C16 and H16 successfully detected the first two DNS tunneling scenarios, they did not at all detect last two DNS exfiltration malware scenarios.

\begin{table}[htb!]
	\centering
	\caption{Summary of Results}
	\begin{tabular}{@{}llll@{}}
		\toprule
		Case Study             & Proposed Method & C16~\cite{cambiaso2016feature} & H16~\cite{homem2016entropy} \\
		\midrule
		Iodine (Web Browsing)            & Yes   & Yes & Yes \\
		Dns2Tcp (File Transfer)          & Yes   & Yes & Yes \\
		FrameworkPOS \\
		(Credit Card Acount Number Theft) & Yes   & No  & No  \\
		Backdoor.Win32.Denis \\
		(Trojan Communication)    & Yes   & No  & No  \\
		\bottomrule
	\end{tabular}
	\label{tbl:results}
\end{table}

\section{Discussion} \label{sec:discussion}
Detecting DNS exfiltration performed by a malware is a difficult challenge. In the results presented, we demonstrate that previous works aimed at detecting DNS tunneling overlooked this issue and are therefore ineffective in the face of this challenge. The main limitation of the methods proposed in previous works lies in both the relatively short data collection phase and the use of per time or per user classification. The results indicate that our proposed method tackles the problem successfully by collecting per domain queries during a predefined window of recent traffic.

Moreover, the per domain approach allows our method to perform an automatic shut down of the data leakage. This automatic blocking can be highly effective for networks in which performing routine manual analysis and investigations of the results may be exhaustive.

\section{Summary and Future Work} \label{sec:summary-future-work}
The ability to detect data leakage in a secure network is crucial and challenging, particularly when leaked over the DNS protocol which is less often monitored and restricted. While the problem has been studied extensively, prior work has largely only focused on a specific class of DNS data leakage, namely DNS tunneling, while leaving the important class of DNS exfiltration malware responsible for leaking millions of stolen credit cards and user credentials in the past decade unaddressed.

In this paper we suggest a method based on anomaly detection, which is capable of detecting both DNS tunneling and DNS exfiltration malware. The proposed method's novelty arises from two points. The first is its inspection of recent history rather than a short time window which may be inefficient for low throughput exfiltration. The second point is our method's inspection based on primary domains, which allows filtering of legitimate services that exchange data over the DNS as well as an automatic blocking of the data leak immediately after detection.

The evaluation of our work is validated on two DNS tunneling tools and two DNS exfiltration malware injected in large-scale DNS traffic. These test subjects are evaluated on our work and compared with two recently published methods in order to demonstrate our method's ability in detecting both classes (DNS tunneling tools and DNS exfiltration malware).

The main limitations of the method are the assumptions of a (1) single and (2) dedicated domain per malware or tunneling tool (see subsection~\ref{subsec:assumptions}). While reasonable, as no malware detected up to the day of writing broke these assumptions, they are not binding. That is, it is possible for a malware to compromise a legitimate domain such that shutting it down using our system will require denying legitimate queries as well as malicious ones. Dealing with this challenge of a data leakage payload that is distributed over queries to different domains and/or uses a compromised domain is proposed for future work.

\ifsubmit
\else
\section*{Acknowledgment}
The authors would like to thank Tal Moran for his wonderful comments on the manuscript. 
\fi

% or


\ifCLASSOPTIONcaptionsoff
  \newpage
\fi



%\bibliographystyle{IEEEtran}
\bibliography{biblio.bib} {}
\bibliographystyle{IEEEtranS}


%\end{IEEEbiography}


%\newpage



%\vfill


\end{document}